\documentclass{jetpl}
\twocolumn

\lat

\title{Magnetic Moments of Negative Parity Baryons from Effective Hamiltonian Approach to
QCD}

\rtitle{Magnetic Moments of Negative Parity Baryons\ldots}

\sodtitle{Magnetic Moments of Negative Parity Baryons from
Effective Hamiltonian Approach}

\author{I.\,M.\,Narodetskii~$^{a,b}$,
M.\,A.\,Trusov~$^{a,}$\/\thanks{e-mail: trusov@itep.ru}}

\rauthor{I.\,M.\,Narodetskii, M.\,A.\,Trusov}

\sodauthor{Narodetskii, Trusov}

\address{$^a$State Research
Center\\Institute of Theoretical and Experimental Physics, \\
Moscow, 117218 Russia
\\~\\
$^b$Moscow Institute of Physics and Technology (State
University),\\Dolgoprudniy, Moscow Region, 141700 Russia }

\dates{9 December 2013}{*}

\abstract{Magnetic moments of $S_{11}(1535)$ and $S_{11}(1650)$
baryons are studied in the framework of the relativistic
three-quark Hamiltonian derived in the Field Correlation Method.
The baryon magnetic moments are expressed via the average kinetic
quark energies which are defined by the fundamental QCD
parameters: the string tension $\sigma$, the current quark masses,
and the strong coupling constant $\alpha_s$.
Resulting magnetic moments for the
$J^P=1/2^-$ nucleons are compared both to model calculations and
to those from lattice QCD.}

\newcommand{\beq}{\begin{eqnarray}}
\newcommand{\eeq}{\end{eqnarray}}
\newcommand{\be}{\begin{equation}}
\newcommand{\ee}{\end{equation}}
\newcommand{\bsl}{\boldsymbol}

\begin{document}

\maketitle

{\bf 1. Introduction.} Magnetic moments encode information about
the leading-order response of a bound system to a soft external
magnetic field. In particular, baryon magnetic moments are
dynamical characteristics which provide valuable insights into
baryon internal structure in terms of quark and gluon degrees of
freedom. In this paper we shall explore the magnetic moments of
negative parity resonances employing the QCD dynamics of a baryon
in the form of the three-quark Effective Hamiltonian (EH). The EH
is derived from the QCD path integral, (see {\it e.g.}
\cite{DDSS}), and was already used in the studies of baryon
spectra without external fields
\cite{NT}~-~\cite{KNV}~\footnote{The extension of the EH to the
case of external magnetic field  has been done recently in Ref.
\cite{S2012}, where the nucleon spectrum as a function of magnetic
field was calculated.}. Within this method the magnetic moments of
the $\frac{1}{2}^+$ octet baryons have been studied analytically
in Ref. \cite{KS}. The model was shown to agree with experiment
within 10$\%$ accuracy. The same accuracy was achieved for the
baryon magnetic moments in Ref. \cite{STW}, where the QCD string
dynamics was investigated from another point of view.

% Negative parity
%partners of the baryon octet arise from excitation of one unit of
%orbital angular momentum.
Although the magnetic moments  of the
$1/2^+$ baryon octet are well-known both experimentally and
theoretically, little is known about their $1/2^-$ counterparts.
Experimentally, magnetic moments of these states can be extracted
through bremsstrahlung processes in photo- and electro-production
of mesons at intermediate energies.  For N(1535) a similar process
$\gamma p\to\gamma\eta p$ can be used \cite{SYVD02},  but to date
no such measurements have been made.

There exist limited number of theoretical studies of the magnetic
moments of negative parity baryons based on  constituent quark
model \cite{SYVD02}, unitarized chiral perturbation theory
(U$\chi$PT)\cite{Hyodo}, chiral constituent quark model ($\chi
CQM$) \cite{STKD12}, Bethe-Salpeter approach (BSA) \cite{CH} and
on the lattice \cite{LA2011}.
% where magnetic moments of the baryon
%resonances have been obtained from the mass shifts.
Comparison
study of magnetic moments for positive- and negative-parity states
offers insight into underlying quark-gluon dynamics. Given that
the mass spectrum of the
%$1/2^+$ and
$1/2^-$ states has been
reasonably well established from the EH, it is instructive to
investigate the magnetic moments of these states. In this paper we
extend the results of Ref. \cite {KS} to the magnetic moments of
the negative parity $S_{11}(1535)$ and $S_{11}(1650)$ resonances.
The paper builds on the previous work presented in Ref. \cite{NSV}
where the employed EH contained the three quark string junction
interaction and the Coulomb potential with the fixed strong
coupling constant.

In Sections 2,3 we briefly discuss the theoretical formalism of EH
method for baryons, including the techniques required to extract
the average quark energies $\omega_i$  which are cornestone of the
present calculation.  As a result one obtains the resonance
magnetic moments without introduction of any fitting parameters.
% and using only string tension, current quark
%masses and the strong coupling constant as an input.
Details of calculation of the magnetic moments for excited
$1/2^{-}$ nucleons and the results are given in Section 4. For the
sake of completeness we also report in this Section the magnetic
moments of the $\frac{1}{2}^+$  octet baryons and $\frac{3}{2}^+$
nucleon resonances. Section 5 contains a summary of the obtained
results.

\noindent{\bf 2. Effective Hamiltonian for Baryons.} The key
ingredient of the EH method is the use of the auxiliary fields
(AF) initially introduced in order to get rid of the square roots
appearing in the relativistic Hamiltonian \cite{P}.  Using the AF
formalism allows one to derive a simple nonrelativistic form of
the EH for the three-quark system which comprises both confinement
and relativistic effects, and contains only universal parameters:
the string tension $\sigma$, the strong coupling constant
$\alpha_s$, and the bare (current) quark masses $m_{a}$,
$a=1,2,3$,
\begin{equation}
\label{eq:H} H=\sum\limits_{a=1}^3\left(\frac
{m_{a}^2}{2\,\omega_a}+ \frac{\omega_a}{2}\right)+H_0+V.
\end{equation}
In Eq. (\ref{eq:H}) $H_0$ is the nonrelativistic kinetic energy
operator for the constant  AF $\omega_a$, the spin-independent
potential $V$ is the sum of the string potential
\begin{equation}\label{eq:string} V_Y({\bf r}_1,\,{\bf r}_2,\,{\bf
r}_3)\,=\,\sigma\,r_{min},\end{equation} with $r_{min}$ being the
minimal string length corresponding to the Y--shaped
configuration, and a Coulomb interaction term
\begin{equation}\label{V_C} V_{\rm C}({\bf r}_1,\,{\bf r}_2,\,{\bf
r}_3)\,=\,-\,C_F\,\sum_{a<b}\,\frac{\alpha_s}{r_{ab}},\end{equation}
arising from the one-gluon exchange. In Eq. (\ref{V_C})
$C_F=\frac{2}{3}$ is the color factor. The constant $\alpha_s$ is
treated either as a fixed parameter, $\alpha_s=0.39$ \cite{NSV} or
as the running coupling constant with the freezing value $\sim
~0.5$ \cite{KNV}. The results for $\omega_a$ for these two cases
coincide with the accuracy better than 1$\%$ (compare Tables 1 and
2 of Refs. \cite{NSV} and \cite{KNV}, respectively). In
what follows we use $\omega_a$ taken from Ref. \cite{NSV}. %We have
\\[2mm]

\noindent{\bf 3. The auxiliary field formalism} The EH
(\ref{eq:H}) depends explicitly on both bare quark masses $m_a$
and the constants AF $\omega_a$ that finally acquire the meaning
of the average kinetic energies of the current quarks \cite{NSV}
\be \omega_a\,=\,\,<\sqrt{{\bsl p}^2_a+m_a^2}>.\ee
% the dynamical quark
%masses.
%These quantities with a good accuracy coincide .
%As the
%first step the eigenvalue problem is solved for each set of
%$\omega_a$; then one has to minimize $\langle H\rangle$ with
%respect to $\omega_a$.
Although being formally simpler the EH  is
equivalent to the relativistic Hamiltonian up to elimination of
AF.

The formalism allows for a very transparent interpretation of
$\omega_a$: starting from bare quark masses $m_a$, we arrive at
the dynamical masses $\omega_a$ that appear due to the interaction
and can be treated as the dynamical masses of constituent quarks.
These have obvious quark model analogs, but are derived directly
using the AF formalism. Due to confinement $\omega_a\,\sim\,
\sqrt{\sigma}\,\sim\,400$ MeV or higher, even for the massless
current quarks.

The baryon mass is given by
\begin{equation}
\label{M_B} M_B\,=\,M_0\,+\,C\,+\,\Delta M_{\rm
string},\end{equation}
\begin{equation}
\label{eq:M_B0}M_0\,=\,\sum\limits_{a=1}^3\left(\frac
{m_{a}^2}{2\omega_a\,}+
\,\frac{\omega_a}{2}\right)\,+\,E_0(\omega_a),\end{equation} where
$E_0(\omega_a)$ is an eigenvalue of the Schr\"{o}dinger operator
$H_0 +V$, and the $\omega_a$ are defined by minimization condition
\begin{equation}\label{eq:mc}
\frac{\partial\,M_0(m_a,\omega_a)}{\partial\,\omega_a}\, =\,0.
\end{equation} The right--hand side of Eq. (\ref{M_B}) contains
the perturbative quark self-energy correction $C$ that is created
by the color magnetic moment of a quark propagating through the
vacuum background field \cite{S2001}. This correction adds an
overall negative constant to the hadron masses. Finally, $\Delta
M_{\rm string}$ in Eq. (\ref{M_B}) is the correction  to the
string junction three-quark potential in a baryon due to the
proper moment of inertia of the QCD string \cite{DNV10}. We stress
that both corrections, C and $\Delta M_{\rm string}$, are added
perturbatively and do not influence the definition of $\omega_a$.

The confinement Hamiltonian contains three parameters: the current
quark masses $m_q$ and $m_s$ and the string tension $\sigma$.  Let
us stress that they are not the fitting parameters. In our
calculations we used $\sigma\,=\,$0.15 GeV$^2$ found in the SU(3)
QCD lattice simulations \cite{Suganuma2003}. We employed the
current light quark masses ${m}_u\,=\,{ m}_d\,=\,9\,$ MeV and the
bare strange quark mass $m_s\,=\,$ 175 MeV.\\[2mm]
\noindent{\bf 4. Magnetic moments of $S_{11}(1535)$ and
$S_{11}(1650)$ resonances.} To calculate the nucleon magnetic
moment one introduces a vector potential ${\bsl A}$ and calculate
the energy shift $\Delta M_B$ due the Hamiltonian ${\cal
H}=H({\bsl A})+H_{\sigma}$ where $H$ is defined by Eq.
(\ref{eq:H}) with the substitution ${\bsl p}_a\to{\bsl
p}_a-e_a{\bsl A}_a$ and \be H_{\sigma}=-\sum_a
\frac{e_a\,{\bsl\sigma_a}}{2\omega_a}{\bsl B},\ee  where $\bsl B$
is an external magnetic field. The magnetic moment operator
consists of contributions from both intrinsic spins of the
constituent quarks that make up the bound state ${\bsl\mu}_S$ and
angular momentum of the three-quark system ${\bsl\mu}_{\bsl L}$
with the center of mass motion removed. Straightforward
calculation using the London gauge  ${\bsl A}=\frac{1}{2}({\bsl
B}\times {\bsl r})$ yields \begin{equation}\label{eq:mm}
{\bsl{\hat\mu}}={\bsl{\hat\mu}}_S+{\bsl{\hat\mu}}_L,\end{equation}
Taking the constituent quarks to be Dirac point particles the spin
contribution in Eq. (\ref{eq:mm})
 is determined by the  effective quark
masses $\omega_i$ \begin{equation} {\bsl\mu}_S\,=\,\sum_a\,
\frac{e_a\,{\bsl\sigma_a}}{2\omega_a},\end{equation} The orbital
contribution in Eq. (\ref{eq:mm}) reads
\begin{equation}\label{eq:mu_L}
{\hat{\bsl\mu}}_L=\sum_a\,\,\frac{e_a}{2\omega_a}\,\,{\bsl
r}_a\times{\bsl p_a}\end{equation} In what follows instead the
usual prescription which is to symmetrize the nucleon wave
function between all three quarks we symmetrize only between
equal-charge (up or down) quarks. In other words for the proton we
use the $uud$ basis in which the $d$ quark is singled out as quark
$3$ but in which the quarks $uu$ are still antisymmetrized. In the
same way, for the neutron we use the basis in which the $u$ quark
is singled out as quark $3$. The $uud$ basis state diagonalizes
the confinement problem with eigenfunctions that correspond to
separate excitations of the quark 3 (${\bsl\rho}$ and
${\bsl\lambda}$\, excitations, respectively). In particular,
excitation of the $\bsl{\lambda}$ variable unlike excitation in
$\bsl{\rho}$ involves the excitation of the ``odd'' quark ($d$ for
$uud$ or $u$ for $ddu$).  The physical P-wave states are not pure
$\rho$ or $\lambda$ excitations but linear combinations of all
states with a given total momentum J. Most physical states are,
however, close to pure $\bsl\rho$ or $\bsl\lambda$ states
\cite{CIK81}.

In terms of the Jacobi variables \be\label{eq:rho}{\bsl\rho}\,=\,
%\sqrt{\frac{\omega_1\omega_2}{\omega_0\,(\omega_1+\omega_2)}}\,
\frac{{\bf r}_1\,-\,{\bf r}_2}{\sqrt{2}}\,,\,\,\,\,\,
{\bsl\lambda}\,=\,
%\sqrt{\frac{\omega_3}{\omega_0\,(m_1+m_2)\,M}}\,
\frac{{\bf r}_1\,+\,{\bf r}_2\, -2\,{\bf r}_3}{\sqrt{6}}\,\ee Eq.
(\ref{eq:mu_L}) reads \beq\label{eq:orbital part}
{\bsl\mu}_L&=&\frac{1}{2}\,(\mu_1+\mu_2)\,{\bsl
l}_{\rho}+\frac{1}{6}\,(\mu_1 +\mu_2+{4\mu_3})\,{\bsl
l}_{\lambda}+\nonumber\\&&+\,\frac{\mu_1-\mu_2}{2\sqrt{3}}({\bsl\rho}\times{\bsl
p}_{\lambda}+{\bsl\lambda}\times{\bsl p}_{\rho}),\eeq where \be
{\bsl l}_{\rho}={\bsl\rho}\times{\bsl p}_{\rho},\,\,\,\,\,{\bsl
l}_{\lambda}={\bsl\lambda}\times{\bsl p}_{\lambda},\ee and where
the quark magnetic moments are expressed in terms of parameters
$\omega_a$ \be\label{eq:quark mm} \mu_a\,=\,\frac{e_a}{2\omega_a}.
\ee Recall that the quantities $\omega_a$  are defined from the
eigenvalues $M(\omega_1,\omega_2,\omega_3)$ of the EH $H$ using
the stationary point equations (\ref{eq:mc}).

Note also that the magnetic moments in Eq. (\ref{eq:quark mm}) are
in quark natural magnetons. To convert it into nuclear magnetons
$\mu_N$, we need to scale the results by the factor
${m_N}/\,{\omega_a}$ . We take $m_N=0.94$ GeV.

The angular operators in (\ref{eq:orbital part}) act on spacial
wave functions $\psi_{1m}^{\rho,\lambda}$ as follows \beq&&l_{\rho
z}\psi_{1m}^{\rho}=m\psi_{1m}^{\rho},\,\,\,\,\,\,l_{\rho
z}\psi_{1m}^{\lambda}=0,\nonumber\\[1mm]&&l_{\lambda
z}\psi_{1m}^{\lambda}=m\psi_{1m}^{\rho},\,\,\,\,\,\,l_{\lambda
z}\psi_{1m}^{\rho}=0,\nonumber\\[1mm]
&&({\bsl\rho}\times{\bsl
p}_{\lambda})_z\psi_{1m}^{\lambda}=m\psi_{1m}^{\rho},\,\,\,\,\,\,({\bsl\rho}\times{\bsl
p}_{\lambda})_z\psi_{1m}^{\rho}=0,\nonumber\\[1mm]
&&({\bsl\lambda}\times{\bsl
p}_{\rho})_z\psi_{1m}^{\rho}=m\psi_{1m}^{\lambda},\,\,\,\,\,\,({\bsl\lambda}\times{\bsl
p}_{\rho})_z\psi_{1m}^{\lambda}=0\label{eq:orbital part 1}\eeq The
contribution of the last term in (\ref{eq:orbital part 1})
vanishes for the pure $\bsl\rho$\,-\,and $\bsl\lambda$\,-
excitations.

By definition, the {\it magnetic moment} $\mu$ of the baryon with
the spin J is the expectation value of the operator $\hat{\mu}^z$
for the state with $M_z=J$ \be
\mu\,=\,<\hat{\mu}^z>\,=\,<JJ|\,\hat{\mu}_{S}^z+\hat{\mu}_L^z\,
%\frac{e_i\,{\bsl\sigma_i}}{2\omega_i}
|\,JJ>\ee In particular, for baryons with
%the total orbital
%momentum
${\bsl L}=0$ where ${\bsl L}$ the angular momentum of the
three-quark system with the correct center of mass motion removed
\beq\label{eq:spin part 1}&&\mu\,=\,\mu_{\rm
spin}^{\frac{1}{2}}\,=\,<\frac{1}{2}^+\frac{1}{2}|\sum_a
\frac{e_a\,{\sigma_{az}}}{2\omega_a}|\,\frac{1}{2}^+\frac{1}{2}>\,=\nonumber\\
&&=\,<\chi^{\lambda}_{\frac{1}{2}\frac{1}{2}}(12;3)\,|\sum_a\,\frac{e_a}{2\omega_a}|
\chi^{\lambda}_{\frac{1}{2}\frac{1}{2}}(12;3)>\,\,
%<\chi^{''}_{\frac{1}{2}\frac{1}{2}}(12,3)\,|\sum_a\mu_a
%\sigma_{az}|\, \chi^{''}_{\frac{1}{2}\frac{1}{2}}(12,3)
%>\,
\nonumber\\&& =\, \frac{1}{3}(2\mu_1+2\mu_2)-\frac{1}{3}\,\mu_3\,
\eeq where $\chi^{\lambda}_{\frac{1}{2}\frac{1}{2}}$ is the
doublet spin function symmetric under interchange
$1{\leftrightarrows 2}$. Eq. (\ref{eq:spin part 1}) is standard
result of the additive quark model for the $\frac{1}{2}^+$ baryons
\cite{Perkins}.

\begin{table*}[t]
\caption{Table 1. The values of $\omega_a$ for the $L=0,1$ baryons
\cite{NSV}.} \label{tab:omega} \centering\vspace{5mm}

\begin{tabular}{ccccc} \hline\hline\\
Baryon&$L$ &Excitation&$\omega_1$ & $\omega_3$\\
\\ \hline\\
$qqq$&0&&0.408&0.408
%&1318&1366&1230&1299&1297
\\
&1&$\rho,\,\lambda$&0.457&0.457

\\\\
\hline \\
$qqs$&$0$&&0.414&0.453
\\
&$1$&$\rho$&0.482&0.459
\\
&$1$ &$\lambda$&0.441 &0.534
\\
\\\hline
\\
$ ssq$&$0$&&0.458&0.419
\\
 & $1$& $\rho$ & 0.520 & 0.424\\
  & $1$& $\lambda$ & 0.483 & 0.506
 \\\\
\hline\hline
\end{tabular}
\end{table*}

In Table 1 we show the parameters $\omega_q$ and $\omega_s$
calculated for the different  baryons with $L=0,1$ in Ref.
\cite{NSV} using the constant value of $\alpha_s\,=\,$0.39. The
symbol q denotes the light quarks $u$ or $d$.  We use the notation
$\omega_1\,=\,\omega_2\,=\,\omega_q$, $\omega_3=\omega_s$  for the
$qqq$ and $qqs$ baryons and $\omega_1\,=\,\omega_2\,=\,\omega_s$,
$\omega_3=\omega_q$  for the $ssq$ baryons. These parameters have
been calculated for the string tension $\sigma=$ 0.15 GeV and the
strong coupling constant $\alpha_s=0.39$ with the values of the
current light quark masses, $m_u\,=\,m_d\,=\,9\,$ MeV,
$m_s\,=\,175$ MeV. The very similar values of $\omega_i$ have been
calculated in Ref. \cite{KNV} where instead the constant
$\alpha_s$ the running coupling constant $\alpha_s(r)$ has been
used with $\alpha_s(\infty)\sim 0.5$.

There is no good theoretical reason why $\omega_q$ need to be the
same in different mesons and baryons. However from the results of
Table 1 we conclude that $\omega_q $ are increased by $\sim 10$
MeV when going from the nucleon to $\Xi$. This variation is
marginal and is within the accuracy of calculations. For ground
states of $\Lambda$ and $\Sigma$ hyperons we obtain
$\omega_q=0.414$ MeV, $\omega_s=0.453$ MeV that agrees with the
corresponding values for the ground state of $K$ meson
\cite{BS2012}.
\begin{table*}[t]\label{tab:mdm}
\caption{Table 2. Baryon magnetic moments of $J^P=\frac{1}{2}^+$
baryons.
%Quark masses from $\cite{NSV}$
}\centering \vspace{5mm}
\begin{tabular}{ccccccccccc}
\hline\hline\\
Baryon &$\omega_q$ & $\omega_s$ & $\mu_u$ & $\mu_d $ & $\mu_s$
&$\mu$&
Expt.\\ \\\hline\hline\\
$p$ &0.408 &&1.53 & -0.77 &&2.29&2.79\\ \\
$n$ & 0.408& &1.53& -0.77& &-1.53& -1.91\\\\
$\Lambda$&0.414&0.453&1.51&-0.76&-0.69 &-0.69&
-0.61\\\\$\Sigma^+$&
  0.414&0.453&1.51&-0.76&-0.69&2.23&2.46\\\\
$\Sigma^0$ &0.414&0.453&1.51& -0.76&-0.69&0.80&0.83\\\\
$\Sigma^-$ &0.414&0.453&1.51&-0.76 & -0.69&-0.91&-1.16\\\\
$\Xi^0$&0.419&0.458&1.485&-0.742&-0.75&-1.40& -1.25\\\\
$\Xi^-$&0.419&0.458&1.689&-0.845&-0.75&-0.50& -0.65\\\\
$\Omega^-$&&0.463&&&-0.671&-2.01&-2.02\\\\
\\\hline\hline
\end{tabular}
\end{table*}
%To some extent the current $s-$quark mass can be probed by the
%experimental hyperon magnetic moments.

The magnetic moments for the $\frac{1}{2}^+$ baryons with $L=0$
are presented in Table 2.
%Note that for the $\Delta$
For the $\frac{3}{2}^+$ baryons one obtains
$\mu_{\Delta^{++}}=3\mu_u=4.575\mu_N$, other moments are $
\mu_{\Delta^+}=2\mu_u+\mu_d=\frac{3}{2}\,\mu_{\Delta^{++}},$
$\mu_{\Delta^0}=0$,
 $\mu_{\Delta^-}=3\mu_d=-\frac{1}{2}\,\mu_{\Delta^{++}}$.
Recall that so far, only the magnetic moment of
$\Delta^{++}(1232)$ has been studied in the reaction $\pi^+p\to
\gamma\pi^+p$ with the result $\mu_{\Delta^{++}} \sim
3.7\,-\,7.5\, \mu_N$  \cite{PDG}. The uncertainty in the number
arises from the ambiguity in the theoretical analysis of the
reaction.

The wave function of an $\frac{1}{2}^-$ resonance is given as a
superposition of two spin ($S = 1/2\,\,\,\, {\rm and}\,\,\,\,
3/2$) states in the $l = 1\,\,$ 70-dimensional representation of
SU(6):
\begin{equation}
|S_{11}(1535)>\,=\,\cos\vartheta\,|^2P_{1/2}>\,-\,\sin\vartheta\,|^4P_{1/2}>\nonumber\end{equation}\begin{equation}
|S_{11}(1650)>\,=\,\sin\vartheta\,|^2P_{1/2}>\,+\,\cos\vartheta\,|^4P_{1/2}>,\nonumber\end{equation}
where mixing angle $\theta$ depends on the hyperfine spin
interaction between the quarks  and the standard spectroscopic
notations $|\,^{2S+1}P_{1/2}>$ are used to indicate the total
quark spin $S = 1/2, 3/2$, orbital angular momentum $L = 1$, and
total angular momentum $J = 1/2$. The corresponding spin-angular
functions are given by \beq &&|\,^2P_{1/2}>\,
=\nonumber\\&&\frac{1}{\sqrt{2}}\left(\sqrt{\frac{2}{3}}\,\,Y_{11}(\bsl\lambda)\,\chi^{\lambda}_{\frac{1}{2}-\frac{1}{2}}-
\sqrt{\frac{1}{3}}\,\,Y_{10}(\bsl\lambda)\,\chi^{\lambda}_{\frac{1}{2}\frac{1}{2}}\right)+\nonumber\\
&&\frac{1}{\sqrt{2}}\left(\sqrt{\frac{2}{3}}\,\,Y_{11}(\bsl\rho)\,\chi^{\rho}_{\frac{1}{2}-\frac{1}{2}}-
\sqrt{\frac{1}{3}}\,\,Y_{10}(\bsl\rho)\,\chi^{\rho}_{\frac{1}{2}\frac{1}{2}}\right),
\eeq where $\chi^{\lambda}_{\frac{1}{2}m_s}$ and
$\chi^{\rho}_{\frac{1}{2}m_s}$ are the two spin functions
symmetric and antisymmetric under interchange $1{\leftrightarrows
2}$, and \beq &&|\,^4P_{1/2}>
%\Psi^{\frac{1}{2}\frac{1}{2}}_{\frac{3}{2}}
=\sqrt{\frac{1}{2}}\,\,Y_{1-1}(\bsl\lambda)\,\chi^s_{\frac{3}{2}\,\frac{3}{2}}-\nonumber\\&&
\sqrt{\frac{1}{3}}\,\,Y_{10}(\bsl\lambda)\,\chi^s_{\frac{3}{2}\,\frac{1}{2}}+
\sqrt{\frac{1}{6}}\,\,Y_{11}(\bsl\lambda)\,\chi^s_{\frac{3}{2}\,-\frac{1}{2}}\eeq

Note that parameters $\omega$ for the $1/2^-$ nucleons depend also
on the type of excitation.
% they are slightly different
%for the ${\bsl\rho}$ and ${\bsl\lambda}$ excitations.
 However,
the difference is marginal and does not exceed 2$\%$, see Table 2
of Ref. \cite{KNV}. In what follows we use the common value
$\omega=0.457$ GeV both for ${\bsl\rho}$ and ${\bsl\lambda}$
excitations.

Straightforward calculation yields~\footnote{The indexes +,\,0
refer to the charge of the nucleon $\frac{1}{2}^-$ states.}
\beq&&\mu(S_{11}^+(1535))=\mu(^2P_{1/2}^+)\cos^2\vartheta+\mu(^4P_{1/2}^+)\sin^2\vartheta\nonumber\\[2mm]
&&-2<^2P_{1/2}^+|\mu_S^z|^4P_{1/2}^+>\sin\vartheta\cos\vartheta=1.24\,\mu_N\eeq
\beq&&\mu(S_{11}^+(1650))=\mu(^2P_{1/2})^+\sin^2\vartheta+\mu(^4P_{1/2})\cos^2\vartheta\nonumber\\[2mm]
&&+2<^2P_{1/2}|\mu_S^z|^4P_{1/2}>\sin\vartheta\cos\vartheta=-0.33\,\mu_N\eeq
\beq&&\mu(S_{11}^0(1535))=\mu(^2P_{1/2}^0)\cos^2\vartheta+\mu(^4P_{1/2}^0)\sin^2\vartheta\nonumber\\[2mm]
&&-2<^2P_{1/2}^0|\mu_S^z|^4P_{1/2}^0>\sin\vartheta\cos\vartheta=\,-0.84\,\mu_N\eeq
\beq&&\mu(S_{11}^0(1650))=\mu(^2P_{1/2}^0)\cos^2\vartheta+\mu(^4P_{1/2}^0)\sin^2\vartheta\nonumber\\[2mm]
&&-2<^2P_{1/2}^0|\mu_S^z|^4P_{1/2}^0>\sin\vartheta\cos\vartheta=\,0.744\,\mu_N\eeq
where
\begin{equation}\label{eq:1}\mu(^2P_{1/2}^+)\,=\,\frac{2}{9}\,\mu_u\,+\,\frac{1}{9}\,\mu_d\,=\,0.23\,\mu_N,\ee
\be\label{eq:2}\mu(^4P_{1/2}^+)\,=\,\mu_u\,+\,\frac{1}{3}\,\mu_d\,=\,1.14\,\mu_N\end{equation}
and
\begin{equation}\label{eq:3} <^4P_{1/2}^+|\mu_z|^2P_{1/2}^+>\,=\,
%-\,<^4P_{1/2}^0|\mu_z|^2P_{1/2}^0>\,=\,
\frac{4}{9}(\mu_u-\mu_d)\,=\,
0.91\,\mu_N\end{equation}
\begin{table*}[t]
\caption{Table 3. Magnetic moments of $J^P=\frac{1}{2}^-$
nucleons. $\omega_q\,=\,0.457\,$ GeV}\centering \vspace{5mm}
\begin{tabular}{ccccccc}
\hline\hline\\
State&${CQM}\,\,\cite{SYVD02}$&${\chi PT}\cite{Hyodo}$&${\chi
CQM}\,\,\cite{STKD12}$&${BSA}\,\cite{CH}$&${LQCD}\,\,\cite{LA2011}$&This
work
\\ \\\hline\hline\\
 $S^+_{11}(1535)$&1.894&1.1&2.085&0.37&-1.8&1.24\\
$S^0_{11}(1535)$&-1.284&-0.25&-1.57&-0.1&-1.0&-0.84\\
 $S^+_{11}(1650)$&0.11&&1.85&&&0.12\\
  $S^0_{11}(1650)$&0.951&&-0.69&&&0.74\\\\
\hline\hline
\end{tabular}
\label{table:NT}
\end{table*}
%\begin{equation}\label{eq:cross term} <^4P_{1/2}^+|\,\mu_z|^2P_{1/2}^+>=\frac{4}{9}(\mu_u-\mu_d)=0.91\,\mu_N\end{equation}
Eqs. (\ref{eq:1}) - (\ref{eq:3}) are written for the positive
charge resonances. For the neutral resonances one should
interchage $\mu_u$ and $\mu_d$ \be
\mu(^2P_{1/2}^0)\,=\,\frac{1}{9}\mu_u+\frac{2}{9}\mu_d\,=\,0,\ee\be
\mu(^4P_{1/2}^0)\,=\,\frac{1}{3}\mu_u+\mu_d\,=\,-0.232\,\mu_N,\end{equation}
and \begin{equation}
<^4P_{1/2}^0|\,\mu_z|^2P_{1/2}^0>=\frac{4}{9}(\mu_d-\mu_u)\,=\,-\,0.926\,\mu_N\end{equation}
Assuming a phenomenological value \cite{IK}
$$\theta\sim -\frac{\pi}{6}$$ we obtain
the results summarized in Table 3. In this Table we also quote the
magnetic moments obtained using other theoretical
models.\\[2mm]
\noindent {\bf 5. Conclusions.} To summarize, we have carried out
a calculation of the magnetic moments of the low-lying negative
parity $S_{11}(1535)$ and $S_{11}(1650)$  resonances. In the
framework of the quark model these resonances are configuration
mixtures of two $SU(6)$ states with excited orbital wavefunctions.
Calculating both the quark spin and orbital angular momentum
contribution for the magnetic moment, the cross terms due to the
configuration mixing, and using the average value of the quark
kinetic energy $\omega=0.457$ GeV obtained from the variational
solution for the einbein field in the EH method we obtain the
values of magnetic moments of the $S_{11}(1535)$ and
$S_{11}(1650)$ listed in Table 3. The results differ from the
magnetic moments of the low-lying $J^P=\frac{1}{2}^-$ nucleon
calculated both in hadronic and quark models and from lattice QCD.
In particular, the lattice results are different, even by sign.
Any future measurement of the magnetic moment would have important
implications in understanding the nature of parity partners of the
nucleon.

Finally we note that the magnetic moments of the other
$\frac{1}{2}^-$ low-lying baryon resonances can similarly be
calculated using EH approach. The
results will be published elsewhere.\\[2mm]

\noindent{\bf Acknowledgements.} The authors thank Yu.~A.~ Simonov
for valuable discussions.

\end{document}